\documentclass{article}

\usepackage{microtype}
\usepackage{graphicx}
\usepackage{amsmath,amssymb,amsthm}
\usepackage{enumitem}
\usepackage{subfigure}
\usepackage{booktabs} 
\usepackage{pifont}
\usepackage{hyperref}

\usepackage[accepted]{mlsys2025edited}

\mlsystitlerunning{Scaling Up Efficient Small Language Models Serving and Deployment for Semantic Job Search }

\begin{document}

\twocolumn[
\mlsystitle{Scaling Up Large Language Models Serving Systems for Semantic Job Search}

\mlsyssetsymbol{equal}{*}

\begin{mlsysauthorlist}
\mlsysauthor{Kayhan Behdin}{li,equal}
\mlsysauthor{Qingquan Song}{ex-li,equal}
\mlsysauthor{Sriram Vasudevan}{li,equal}
\mlsysauthor{Jian Sheng}{li,equal}
\mlsysauthor{Xiaojing Ma}{li,equal}
\mlsysauthor{Z Zhou}{li,equal}
\mlsysauthor{Chuanrui Zhu}{li}
\mlsysauthor{Guoyao Li}{li}
\mlsysauthor{Chanh Nguyen}{li}
\mlsysauthor{Sayan Ghosh}{li}
\mlsysauthor{Hejian Sang}{li}
\mlsysauthor{Ata Fatahi Baarzi}{li}
\mlsysauthor{Sundara Raman Ramachandran}{li}
\mlsysauthor{Xiaoqing Wang}{li}
\mlsysauthor{Qing Lan}{li}
\mlsysauthor{Vinay Y S}{li}
\mlsysauthor{Qi Guo}{li}
\mlsysauthor{Caleb Johnson}{li}
\mlsysauthor{Zhipeng Wang}{li}
\mlsysauthor{Fedor Borisyuk}{li}

\end{mlsysauthorlist}

\mlsysaffiliation{li}{LinkedIn}
\mlsysaffiliation{ex-li}{Work done at LinkedIn}

\mlsyscorrespondingauthor{Zhipeng Wang}{zhipwang@linkedin.com}

\mlsyskeywords{Machine Learning, MLSys}

\vskip 0.3in

\begin{abstract}
Large Language Models (LLMs) have demonstrated impressive quality when applied to predictive tasks such as relevance ranking and semantic search. However, deployment of such LLMs remains prohibitively expensive for industry applications with strict latency and throughput requirements. In this work, we present lessons and efficiency insights from developing a purely text-based decoder-only Small Language Model (SLM) for a semantic search application at LinkedIn. Particularly, we discuss model compression techniques such as pruning that allow us to reduce the model size by up to $40\%$ while maintaining the accuracy. Additionally, we present context compression techniques that allow us to reduce the input context length by up to $10$x with minimal loss of accuracy. Finally, we present practical lessons from optimizing the serving infrastructure for deploying such a system on GPUs at scale, serving millions of requests per second. Taken together, this allows us to increase our system's throughput by $10$x in a real-world deployment, while meeting our quality bar.
\end{abstract}
]

\printAffiliationsAndNotice{\mlsysEqualContribution}

\section{Introduction}

Large Language Models (LLMs) have demonstrated impressive predictive performance in a variety of industrial tasks. Among such tasks, LLMs have been particularly successful when applied to relevance ranking for semantic search applications, where the goal is to determine how relevant an item is to a user's query. Recent work~\citep{relevance1,relevance2,relevance3,relevance4} has shown LLMs enjoy strong semantic understanding, allowing them to be powerful relevance judges. Despite their impressive performance, serving such LLMs in an industrial setting remains challenging. This is due to the fact that LLMs can introduce unacceptable latency to real-world systems, or require significant compute resources to meet the throughput goals of the system. Therefore, improving the efficiency of LLMs for ranking and relevance remains of high interest.

In this work, we study a real-world semantic search system from an efficiency perspective. At the heart of this system, we utilize a text-based decoder-only Small Language Model (SLM) that scores the relevance of items and queries. In particular, we discuss the techniques used to improve the efficiency of this ranker model to meet our strict latency and throughput requirements, while ensuring the quality remains high. We present, in details, the lessons we learned from deploying our SLM-powered semantic search system to a large-scale professional social network.

To be more specific, we explore several optimization and compression techniques, with the goal of improving the system's throughput. In particular, we explore:
\begin{enumerate}
    \item \textbf{Model Compression via Pruning}: Pruning is a model compression technique to obtain smaller models, by removing certain redundant structures from the model. We use model pruning to obtain high-quality smaller models that enjoy higher inference throughput.
    \item \textbf{Item Description Summarization}: Given that the complexity of the self-attention mechanism scales quadratically with the input context length, we design novel summarization techniques, where we create summaries of item descriptions offline using another language model trained with reinforcement learning. These summaries are then used by the SLM ranker in the online pipeline, allowing for faster online inference.
    \item \textbf{Serving Infrastructure Optimization}: Our semantic search workload operates in a regime that is different from common LLM use cases, for example, in chatbots. As we discuss, we serve a prefill-only use case, with extremely high rate of Requests Per Second (RPS). We share several lessons from optimizing our online and streaming serving stack, including lessons from optimizing the SGLang~\citep{sglang} inference engine for high RPS prefill-only workloads. 
\end{enumerate}

\subsection{Related Work}

Several papers have studied compressing foundation models into smaller ones; Examples include~\citet{pruning-related1,pruning-related2} which study model compression techniques such as pruning and knowledge distillation for foundation models. In another work,~\citet{pruning-related3} explore several techniques such as quantization and distillation to improve the efficiency of a foundation model. We also refer to~\citet{pruning-related4} for a survey of model compression techniques. 

Lengthy input to language models reduces inference speed, increases memory requirements and also negatively impacts user experiences, especially in a real-world ranking system. Multiple efficient methods have been proposed for prompt compression such as Nano-Capsulator \citep{chuang2024learning} and Gisting \citep{mu2023learning}. \cite{li2024prompt} provides a comprehensive survey of prompt compression for large language models. 

We note that, however, our work is different from the aforementioned ones. In our work we focus on an SLM used in a semantic search system, rather than a general foundation model. In particular, there has been a limited exploration of model compression techniques for relevance ranking applications. In addition, our context compression needs to work for a vast amount of versatile item descriptions, while keeping natural language format for interpretability. We share our learnings from applying model and context compression techniques, as well as GPU deployment lessons.

There are several notable difference between our work and the previous work studying LLMs for relevance in industrial applications.~\citet{related-pinterest} study using an LLM as a relevance judge for semantic search at Pinterest. However, instead of deploying the LLM online, they distill it to a smaller feed-forward neural network, due to the difficulty of serving LLMs at scale. In contrast, as a result of our optimizations, we directly deploy an SLM online, which as we discussed, comes with practical challenges. This further signifies the importance of the techniques we present in this paper. Similarly,~\citet{related-ebay} discuss distilling an LLM relevance judge into a non-LLM architecture.
In another example,~\citet{related-walmart} discuss distilling an LLM into a BERT-style model for a relevance use case. This is in contrast to our work where we use a decoder-only model. Additionally, certain features such as item description are dropped from their online serving stack due to efficiency reasons, while we implement summarization techniques that allow us to include such long context features in our online system. This is another indication of the importance of the context compression methods we develop.

\section{Semantic Job Search}\label{sec:sjs-intro}
LinkedIn's Semantic Job Search introduces an AI-driven approach to job discovery by shifting from keyword matching to intent understanding. The system leverages natural language input to capture user goals and preferences, reducing reliance on exact lexical overlap between queries and job postings. This semantic representation enables more robust retrieval, addressing vocabulary mismatches while offering a search experience that better reflects how members naturally describe career aspirations.

\begin{figure*}[ht]
    \centering
    \includegraphics[width=0.7\linewidth]{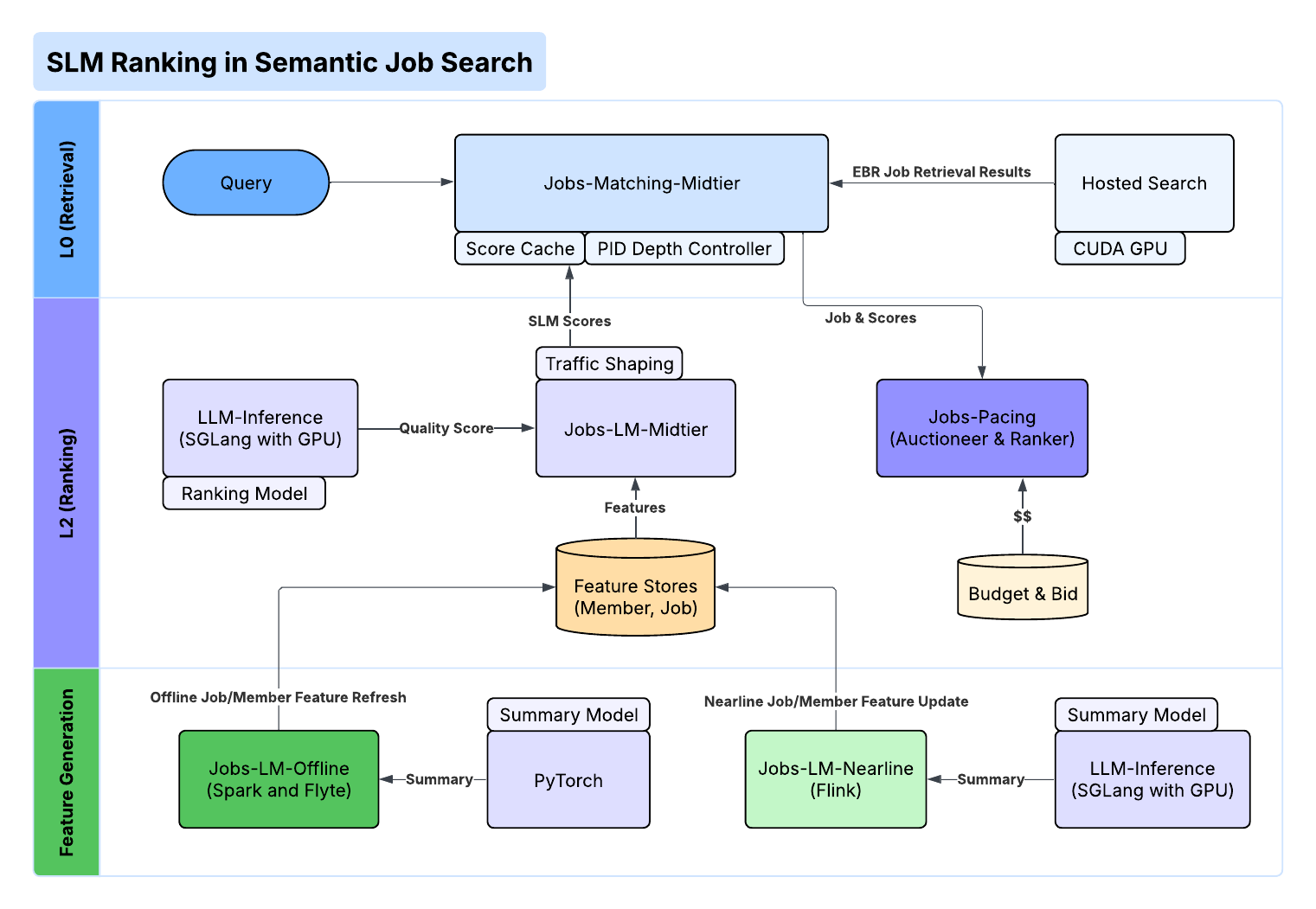}
    \vspace{-20pt}
    \caption{Overall pipeline of LinkedIn's Semantic Job Search. }
    \label{fig:overview}
\end{figure*}

The retrieval layer begins with a query understanding service that parses the user’s input, generates a query embedding, and runs embedding-based retrieval (EBR) on CUDA-accelerated GPUs with exhaustive vector search \citep{10.1145/3705328.3748116} to collect a broad pool of candidate jobs. In the ranking layer, these candidates are refined by a Cross-Encoder SLM hosted on the SGLang, which integrates search query with job and member features to produce quality scores for final ranking. To keep the system efficient and scalable, the ranking layer incorporates score caching, a ranking depth controller to regulate how many candidates proceed through deeper ranking, and traffic shaping to smooth request bursts—all of which optimize load management and improve search outcomes. The features and summarized job descriptions used by the SLM are pre-computed in a hybrid pipeline: an offline system with Spark and Flyte for large-scale batch inference, and a nearline system with Flink for low-latency updates. These are persisted in a distributed storage system, and fetched at request time with very low latency. Finally, the auction layer applies budget and pacing logic to balance user relevance, engagement, and business objectives, ensuring both member satisfaction and marketplace health with balanced recall and precision. An overview of the overall system is shown in Figure~\ref{fig:overview}.

 A product policy specifies how to grade the relevance of each (query, job) pair. Concretely, we train a 
 7B LLM
 to generate 5-point graded scores along several dimensions, together with rationales, such that it aligns with the product policy. These dimension-wise grades are aggregated into a final label for the pair. The retrieval and ranking systems are then distilled from this teacher so that the end-to-end stack optimizes the Normalized Discounted Cumulative Gain at the tenth position (NDCG@10) using teacher-provided labels.

In this work, we focus on the ranking layer of the system. The ranking stage employs an SLM to score the relevance of a user's search query $q$ to each retrieved job $i$. This SLM is a decoder-only model. We denote the structured job attributes - title, company, location, employment type, and remote-work eligibility - by $\text{metadata}_i$, and the free-text job description by $\text{desc}_i$.
For each ($q$, $i$) pair, we construct a prompt
\begin{equation}\label{eq:prompt}
    \text{prompt}(q,i)=\text{system prefix},q,\text{metadata}_i,\text{desc}_i,\text{suffix},
\end{equation}
where $\text{system prefix}/\text{suffix}$ include chat-template tags and instructions to decide whether the item matches the query. Passing this prompt through the decoder yields logits for the next token. Let $\text{logit}_{\text{yes}},\text{logit}_{\text{no}}$ denote the logits for the tokens ``yes'' and ``no'' corresponding to the last input token, respectively. Following prior work~\citep{related-pinterest,qwen3emb}, we compute
\begin{equation}\label{eq:rel-score}
    (p_{\text{yes}},p_{\text{no}})=\text{Softmax}(\text{logit}_{\text{yes}},\text{logit}_{\text{no}})
\end{equation}
which yields probabilities used to rank items by relevance.

SLM Training uses a multi-stage pipeline. First, we distill the 7B teacher into a 0.6B model to obtain a smaller model that generates graded labels with rationales. Next, we map the teacher's ordinal grades to $(p_{\text{yes}}^*,p_{\text{no}}^*)$ ``soft labels,'' and perform Supervised Fine-Tuning (SFT) to minimize the Kullback–Leibler (KL) divergence between the teacher’s soft labels and the SLM’s predictions, turning the SLM from a reasoning model into a binary classifier:
\begin{equation}\label{eq:kl_loss}
    \text{loss}=KL((p^*_{\text{yes}},p^*_{\text{no}})\Vert(p_{\text{yes}},p_{\text{no}}))
\end{equation}
with $\mathrm{KL}(p\Vert q)=\sum_i p_i\log\!\left(p_i/q_i\right)$.

We train on 200k examples for up to 5 epochs using the prompt in~\eqref{eq:prompt}. The prompt is dominated by $\text{desc}_i$, whose length varies widely (median $\sim\!900$ tokens; max $>\!2100$). We truncate $\text{desc}_i$ so that the total prompt length does not exceed 2048 tokens during both training and inference.

Evaluation uses a holdout set labeled by the teacher. The SLM ranks candidates by $p_{\text{yes}}$, and we report NDCG@10.

To meet our job search traffic across multiple product facets, our ranking system has to be able to score 3.15M items per second, making the efficiency of the ranker of utmost importance. 

\section{Discussion of compression techniques}

\subsection{Model compression via structured pruning}
To improve the throughput, we explore model compression via pruning. Model pruning~\citep{obd,obs,benbaki2023fast} is a model compression technique where some model weights or components are removed to obtain smaller models, with a smaller computational footprint. Model pruning has been successfully applied to LLMs~\citep{frantar2023sparsegpt,sun2023simple,meng2024alps}, allowing model compression with a small loss of quality. In this work, as we are interested in deploying our model on GPUs, we focus on structured pruning. In structured pruning, certain model components (neurons, layers, attention heads) are removed from the model, resulting in models that are smaller with fewer parameters. Such pruned models can then be deployed without requiring specialized hardware or kernels, leading to improved throughput and latency. This is in contrast to unstructured pruning, where inference acceleration might only be possible on certain hardware. Specifically, in our work, we consider pruning of:
\begin{itemize}[noitemsep,topsep=0pt,parsep=0.5pt,partopsep=0pt]
    \item Hidden neurons in Feed-Forward MLP layers
    \item Whole transformer blocks.
\end{itemize}
We use OSSCAR~\citep{meng2024osscar} for pruning hidden neurons in MLP blocks. We also experiment with removing complete transformer blocks in the model---we present our results in Section~\ref{subsec:pruning-exp}. These pruning approaches result in smaller models that can be accelerated on any hardware. In particular, removing hidden neurons in MLP layers reduces model's intermediate size, while removing transformer blocks results in a model with a smaller number of hidden layers. After performing pruning, we fine-tune the pruned model to recover any accuracy lost due to compression.

\subsection{Context compression via summarization}

\begin{figure}[ht]
    \centering
    \includegraphics[width=1.0\linewidth]{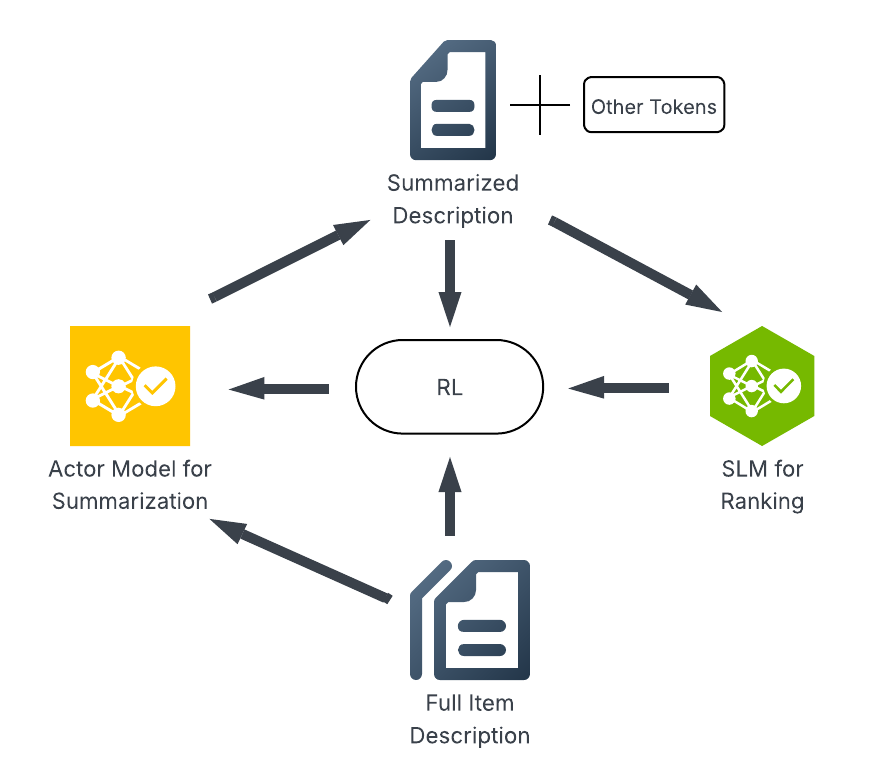}
    \vspace{-20pt}
    \caption{\textbf{Overall pipeline of context summarization with RL.} We start with full item description, which is processed by the actor model to generate a summarized version. The summarized description is then combined with other tokens such as metadata before feeding into the SLM. The reward to the RL is calculated using the output of the SLM, coupled with the length of full and summarized description to update the parameters of the actor model. The parameters of the SLM are kept frozen throughout the process.}
    \label{fig:flow}
\end{figure}

Large Language Models (LLMs) have demonstrated impressive capabilities in contextual learning, leverage more complex and sophisticated input to achieve superior performances. However, longer prompt length leads to higher latency and larger memory consumptions. In particular, in our use case of SLM, the time-to-first-token (TTFT) latency increases quadratically with regard to input length, limiting its applicability to serve real-time traffic for billions of users.

We explore context compression techniques to improve further latency and throughput, while maintaining model quality as much as possible. Prompt / context compression has gained significant research interest \citep{li2024prompt} in recent years. For example, Gisting \citep{mu2023learning} trains an LM to compress prompts into smaller sets of "gist" tokens which can be cached and reused during inference. However, this is usually used for general system prompt / task descriptions, and cannot be generalized to a vast amount of heterogeneous item descriptions. Nano-Capsulator \citep{chuang2024learning} compresses original prompts into natural language formatted prompt while maintaining the prompt utility and transferability. It uses a semantics preserving loss coupled with a reward function featuring length constraints. In our use case, text descriptions can be verbose and contain information that isn't useful for ranking, which cannot be distinguished with semantic loss alone.

We propose to fine tune a language model for generating summarizations based on raw context with Reinforcement Learning (RL). Given our two objectives on 1) shorten context length and 2) maintain model quality, our reward signal is constructed with two components respectively: 1) an output sequence length penalty and 2) KL divergence loss between the SLM outputs $(p_{\text{yes}},p_{\text{no}})$ on summarized context and raw context. Equation \eqref{eq:reward} shows the reward signal, where the subscript denotes summarized and raw context respectively, and $w$ is a hyper-parameter controlling the relative importance from length penalty. 

\begin{equation}\label{eq:reward}
\text{reward} = -KL(p_\text{sum}\Vert p_\text{raw}) - w  (\frac{len_\text{sum}}{len_\text{raw}})^2.
\end{equation}

Note that the approach we proposed here shares similarities with reinforcement learning from verifiable rewards \citep{guo2025deepseek, lambert2024tulu}, and recent work on reinforcement learning from performance / user feedback \citep{jiang2025improving, han2025reinforcement}. The diagram in Figure~\ref{fig:flow} is an overview of the whole pipeline.

Our RL training pipeline is built on top of verl \citep{sheng2024hybridflow}, and we use GSPO \citep{zheng2025group} as the optimization algorithm. We present experiment results in Section \ref{subsec:compression-exp}, where we show that our compression method can achieve over 10X context compression ratio with only 2\% quality drop. We also include comprehensive ablation studies regarding the necessity of reinforcement learning and length penalty design.

\section{Experimental Results}

\subsection{Pruning Experiments}\label{subsec:pruning-exp}
In this section, we present our model pruning experiments and the insights from our experiment. The experiments in this section use prompts with full item description (that is, no summarization is not applied in this section). In the next sections, we discuss integrating summarization and pruning.

We start by studying the effect of pruning hidden neurons in the MLP layers of the model. To this end, we start from a 0.6B model that has been fine-tuned on our relevance data. Next, we remove 50\% of hidden neurons in all MLP layers via OSSCAR. Pruning methods such as OSSCAR often use a small dataset (known as the calibration data) to choose which neurons, for example, can be removed from the model. Following the prior work~\citep{behdin2025efficient}, we use in-domain data as for calibration data, using around 40M tokens from our training prompts. Additionally, after pruning, we perform SFT to further recover the accuracy lost after pruning. The results for pruning 50\% of hidden neurons in MLP layers is presented in Table~\ref{table-osscar}.

\begin{table}[t]
\caption{The results for pruning hidden neurons in MLP layer via OSSCAR, with and without SFT after pruning. The pruned model has around 460M parameters, compared to the the unpruned version with 600M.}
\label{table-osscar}
\vskip 0.15in
\begin{center}
\begin{small}
\begin{sc}
\begin{tabular}{lc}
\toprule
Model & NDCG@10 Change \\
\midrule
Unpruned & - \\
50\% MLP Pruning & $-0.0095$ \\
50\% MLP Pruning + SFT & $-0.0046$ \\
\bottomrule
\end{tabular}
\end{sc}
\end{small}
\end{center}
\vskip -0.1in
\end{table}

We observe that even when pruning half of the hidden neurons in the MLP layers, we observe a modest drop in accuracy (less than $1\%$ in terms of NDCG@10). We also observe that most of this accuracy drop is recovered after performing SFT on the pruned model. After pruning, we obtain a model with around 460M, a 25\% reduction in model size.

Next, we explore how to remove whole transformer blocks from the model. To this end, we remove one transformer block at a time, and then evaluate the resulting model's quality. We show the results in Table~\ref{table-layer-sensitivity}. We observe that the model quality seems to be least sensitive to the removal of last few layers, while the first and middle layers seem to be most important and need to be preserved.

\begin{table}[t]
\caption{Results for pruning transformer blocks from the model. We remove a transformer block at a time, and evaluate the model's quality. We see that removing last layers has a smaller effect on model's quality.}
\label{table-layer-sensitivity}
\vskip 0.15in
\begin{center}
\begin{small}
\begin{sc}
\begin{tabular}{lc}
\toprule
Layer to remove & NDCG@10 Change  \\
\midrule
No Layer Removed & -\\
First Layer & $-0.3356$\\
Second Layer & $-0.0296$ \\
Third Layer & $-0.0133$\\
Tenth Layer &  $-0.0166$\\
Two Before Last Layer & $-0.0001$\\
One Before Last Layer & $-0.0004$\\
Last Layer & $-0.0009$\\
\bottomrule
\end{tabular}
\end{sc}
\end{small}
\end{center}
\vskip -0.1in
\end{table}

Combining our experiments from Tables~\ref{table-osscar} and~\ref{table-layer-sensitivity}, we combine MLP pruning and transformer layer pruning into one pipeline, to obtain even smaller models. To this end, we first follow the same recipe from Table~\ref{table-osscar} (with SFT). Then, we remove the last few transformer layers from the model to obtain a smaller one. Next, we run SFT on top of this model that has undergone two rounds of pruning. The results for this case are presented in Table~\ref{table-layer-osscar}. Based on our results, we are able to reduce the model size by 45\% while losing less than $1\%$ of the quality, allowing us to significantly increase our inference throughput with a marginal quality loss.

\begin{table}[t]
\caption{Integrating MLP and transformer block pruning. We consider 50\% pruning of MLP hidden neurons, and removing last few transformer blocks of the model. We include SFT steps after each pruning step. }
\label{table-layer-osscar}
\vskip 0.15in
\begin{center}
\begin{small}
\begin{sc}
\begin{tabular}{lcc}
\toprule
Model Size &  NDCG@10 Change  \\
\midrule
600M &  - \\
375M (50\% MLP + 8 layers)  & $-0.0079$ \\
350M (50\% MLP + 10 layers)  & $-0.0074$ \\
330M (50\% MLP + 12 layers)  & $-0.0080$ \\

\bottomrule
\end{tabular}
\end{sc}
\end{small}
\end{center}
\vskip -0.1in
\end{table}

\subsection{Context Compression Experiments}\label{subsec:compression-exp}

As mentioned in Section~\ref{sec:sjs-intro}, complete item descriptions have a median length of 900 tokens and a maximum of over 2300. This results in over $94\%$ of the SLM prompt being comprised of just the description, and around $10\%$ of all prompts being truncated due to the 2048 token budget. Excluding the description entirely results in a major performance drop (Table~\ref{table-summarization-prompt}), indicating that there is significant value in retaining this field for the relevance prediction task.

We therefore utilize a more powerful 1.7B LLM to summarize these descriptions, since item-specific inference can be precomputed offline and updated using streaming solutions.

\begin{table}[t]
\caption{Context compression with prompt engineering.}
\label{table-summarization-prompt}
\vskip 0.15in
\begin{center}
\begin{small}
\begin{sc}
\begin{tabular}{lcc}
\toprule
Approach & NDCG@10 & Compression \\
\midrule
No item description & $-9\%$ & $-100\%$ \\
Stop word removal & $-0.5\%$ & $-16\%$ \\
``Summarize'' prompt & $-2.5\%$ & $-52\%$ \\
YAML prompt & $-2.7\%$ & $-74\%$ \\
``Key phrases'' prompt & $-1.5\%$ & $-61\%$ \\
\bottomrule
\end{tabular}
\end{sc}
\end{small}
\end{center}
\vskip -0.1in
\end{table}

\subsubsection{Prompt Engineering}
Our preliminary experiments combined heuristics with a pure prompt-engineering approach, using the open-weights version of this model:
\begin{itemize}[topsep=0pt,parsep=0pt,partopsep=0pt]
    \item \textbf{Stop word removal:} We used heuristics to strip tokens that matched those in a special dictionary since we observed that the 1.7B model was unable to follow a ``remove stop words'' instruction consistently.
    \item \textbf{``Summarize this job description'' prompt:} We prompted the model to directly output summaries, but observed shorter responses that didn't capture the job requirements very well.
    \item \textbf{YAML-focused prompt:} We used the model to extract facts from the job description and present it as a structured output. This worked well from an instruction-following perspective, but we observed similar results as in the previous prompt scenario.
    \item \textbf{``Rewrite using key phrases that job seekers might search for'' prompt:} By avoiding the ``summarize'' keyword, the model generates longer descriptions, resulting in a smaller performance drop.
\end{itemize}

Our results are captured in Table~\ref{table-summarization-prompt}. We see that a pure prompt-oriented approach is insufficient to achieve significant length compression without considerable model performance degradation.

Some additional avenues to explore in the prompt-engineering space include using DSPy \cite{khattab2024dspy} to optimize the summary prompt based on rewards and constraints.

\subsubsection{Reinforcement Learning}
A fundamental issue with the prompt-engineering approach is that the generated summary is not aligned with the job search SLM, leading to a performance degradation. Furthermore, setting limits on the summary lengths is also error-prone with just a prompt-based approach. We therefore leverage Reinforcement Learning based approaches to tune the weights of the summary model (the actor) instead, aligning it with the SLM (reward model) while also incorporating a length penalty. RL results in a more generalized summary model compared to a Supervised Fine-Tuning approach. Furthermore, we don't need to provide example summaries in our training dataset. We reuse the simple ``Summarize'' prompt as the input to the actor.

For the RL algorithm, we experimented with Group Relative Policy Optimization (GRPO)~\cite{shao2024deepseekmath}, GRPO Done Right (Dr. GRPO, an improvement over GRPO)~\cite{liu2025understanding}, as well as the more sample-efficient Group Sequence Policy Optimization (GSPO)~\cite{zheng2025group}.

We also experimented with two different length penalty formulations described in Equations~\eqref{eq:summ-v1} and~\eqref{eq:summ-v2},
\begin{equation}\label{eq:summ-v1}
P_1(L_o,L_c)=
\begin{cases}
0, &
\begin{aligned}[t]
&\text{if } L_o<m\\
&\text{or } r \le \tau,
\end{aligned}
\\[8pt]
-w \left(\dfrac{r-\tau}{1-\tau}\right)^{2}, & \text{otherwise,}
\end{cases}
\end{equation}
\begin{equation}\label{eq:summ-v2}
P_2(L_o,L_c)= -w r^{2}
\end{equation}
where $L_o$ and $L_c$ are the original and compressed token lengths respectively, $r$ is the compression ratio ${L_c}/{L_o}$, $m$ is the maximum acceptable length and $\tau$ is the largest acceptable compression ratio.

The motivation for $P_1$ is to encourage compression only for  descriptions longer than $m$, and to keep the compression ratio $r=L_c/L_o$ at or below a target $\tau$ for most cases. Between $\tau L_o$ and $L_o$, the penalty increases smoothly, using a quadratic term so that longer summaries incur disproportionately larger costs. By contrast, $P_2$ penalizes descriptions of all lengths, biasing the actor toward shorter outputs at potential expense to SLM performance. We adopt a quadratic norm rather than a linear one because it minimizes the expected overshoot $\mathbb{E}\!\left[(r-\tau)_{+}\right]$, whereas a linear form primarily reduces the tail probability $Pr(r > \tau)$. From a throughput standpoint, the magnitude of violations above $\tau$ matter more than their mere frequency, making the quadratic choice preferable.

In our experiments, we fixed $m=256,~\tau=1/3$ based on system requirements and only tune $w$ for various algorithm-penalty combinations. Our results are captured in Table~\ref{table-rl-experiments}. An interesting observation with $P_1$ is that as we increase $w$, we see more than $66\%$ compression despite $\tau=1/3$ because the asymmetric reward creates a \textit{margin-seeking} or risk-averse effect. When $w$ is large, any sample that exceeds $\tau$ is considerably worse than those that are below it, so the policy learns to hedge by moving well below $\tau$. With $P_2$, we observe that it lags in performance compared to $P_1$ for less aggressive compression (a $2\%$ NDCG drop for  $89\%$ compression, compared to $< 2\%$ for a similar compression using $P_1$). However, as we attempt to compress more aggressively (eg: $94\%$), $P_2$ performs better, likely because we don't zero out the penalty as the summaries get shorter, so the gradient updates are not sparse and the model still learns. Finally, coming to GSPO, we observe that it is able to deliver a better performance-compression tradeoff with $P_2$, as compared to Dr. GRPO.

With a maximum acceptable drop of $2\%$ NDCG@10, we therefore chose GSPO with the $P_2$ length penalty and $w=0.4$, giving us a $93\%$ reduction in p50 and p99 item description lengths. Comparing this to the ``no description'' case from Table~\ref{table-summarization-prompt}, we see that we can compress context worth $7\%$ NDCG@10 into descriptions that are just $7\%$ of their original lengths. The original item descriptions make up almost $94\%$ of the SLM's prompt. This drops to $54\%$ with the compressed item descriptions.

Beyond the $93\%-95\%$ compression range, we almost always notice a precipitous drop in model performance. At these high compression rates, we believe this is due to the summary model being forced to drop information, since it can no longer capture this using the language's vocabulary. Updating the algorithm to incentivize outputs in a more token-efficient language might alleviate this challenge. Another observation is that the summary model learns to generate well-formed text, perhaps due to the reward model having been trained on complete job descriptions. We see a terse note-taking style emerge only when the penalty weight is high. Finally, since the reward model was trained on English job descriptions and the summary prompt is also in English, we observe a tendency for the trained actor to output summaries in English even when the original job description is in another language. The summaries are still valid though, due to the LLM's inherent translation capabilities.

\begin{table*}[t]
\caption{RL Experiments. We report the p99 compression here, although we observed fairly similar compression at p50 too.}
\label{table-rl-experiments}
\vskip 0.15in
\begin{center}
\begin{small}
\begin{sc}
\begin{tabular}{lllcc}
\toprule
Algorithm  & Length Penalty & Penalty Weight & NDCG@10 & Compression \\
\midrule
``Summarize'' prompt & - & - & $-2.5\%$ & $-52\%$ \\
\midrule
GRPO & $P_1$ & $w=0.0$ & $-1.8\%$ & $-52\%$ \\
\midrule
Dr. GRPO & $P_1$ & $w=0.0$ & $-1.0\%$ & $-66\%$ \\
Dr. GRPO & $P_1$ & $w=0.05$ & $-1.1\%$ & $-69\%$ \\
Dr. GRPO & $P_1$ & $w=0.1$ & $-1.6\%$ & $-84\%$ \\
Dr. GRPO & $P_1$ & $w=1.0$ & $-2.0\%$ & $-92\%$ \\
Dr. GRPO & $P_1$ & $w=10.0$ & $-2.4\%$ & $-94\%$ \\
\midrule
Dr. GRPO & $P_2$ & $w=0.1$ & $-2.0\%$ & $-89\%$ \\
Dr. GRPO & $P_2$ & $w=0.5$ & $-2.2\%$ & $-94\%$ \\
Dr. GRPO & $P_2$ & $w=1.0$ & $-2.6\%$ & $-95\%$ \\
Dr. GRPO & $P_2$ & $w=10.0$ & $-3.9\%$ & $-98\%$ \\
\midrule
GSPO & $P_2$ & $w=0.3$ & $-1.7\%$ & $-91\%$ \\
\textbf{GSPO} & \boldmath $P_2$ & \boldmath $w=0.4$ & \boldmath $-2\%$ & \boldmath $-93\%$  \\
GSPO & $P_2$ & $w=1.0$ & $-2.2\%$ & $-95\%$ \\
\bottomrule
\end{tabular}
\end{sc}
\end{small}
\end{center}
\vskip -0.1in
\end{table*}

\subsection{Integrating Model and Context Compression}
We combine model and context compression experiments discussed above in our final pipeline. From our experimental results, we MLP pruning via OSSCAR, and removing the last few layers of the model (Table~\ref{table-layer-osscar}). In our final pipeline, we prune 50\% of the MLP hidden neurons, as well as the last 8 transformer blocks, leading to a 375M model. After pruning, we perform SFT, either using the full item descriptions or summarized item descriptions. We then evaluate the model on the evaluation data that only includes the summarized item descriptions (which is the production setup). The overall quality results are presented in Table~\ref{table-prune-summary}. We see that after pruning and summarization, the model quality drops by less than 2\%, while as we discuss, taken together, these compression techniques allow us to significantly improve the ranking throughput. Interestingly, we observe that performing SFT after pruning using the summarized data results in a marginally higher NDCG value.

\begin{table}[t]

\caption{Integrating model and context compression. The first row does not include any compression. ``SFT on summarized'' indicates that if the fine-tuning after pruning data uses the summarized item descriptions or not. Similarly, ``Evaluation on summarized'' indicates that if the evaluation data uses the summarized item descriptions or not. In production, we use the pruned model and summarized item descriptions for SFT and evaluation (the last row).}
\label{table-prune-summary}
\vskip 0.15in
\begin{center}
\begin{scriptsize}
\begin{sc}
\begin{tabular}{lcccc}
\toprule
Model Size & SFT on  & Evaluation on  &  NDCG@10 \\
& summarized & summarized & \\
\midrule
600M & \ding{55} & \ding{55} & 0.8950 \\
375M & \ding{55} & \ding{51} & 0.8786\\
375M & \ding{51} & \ding{51} & 0.8788\\

\bottomrule
\end{tabular}
\end{sc}
\end{scriptsize}
\end{center}
\vskip -0.1in
\end{table}

\section{Inference Benchmarking and Optimization}
In this section, we present our results regarding the inference and deployment of our SLM in the semantic search application.
\subsection{Offline Benchmarking}
Before moving to an online serving setup, we perform extensive offline benchmarking to understand how model and context compression allow us to increase the system throughput. In particular, we start by benchmarking the raw inference engine, to demonstrate the benefits of our compression techniques. 

To this end, we use SGLang (version 0.4.6) inference engine with NVIDIA H100 GPUs for inference (CUDA version 12.6). To simulate the online traffic, for each query we create 10 different prompts using 10 different candidate items. We then send the batch of prompts to the engine at random times for scoring, according to a Poisson process. We call each batch of 10 prompts a request.  

We increase the average Requests Per Second (RPS), until the 95-th percentile end-to-end latency reaches 500 milliseconds, which is our acceptable latency. This allows us to obtain the maximum traffic throughput the engine can handle. 

We consider several setups to benchmark. First, we use the (unpruned) 0.6B model, using full item description without summarization. We use our training data to form the input to the engine. This leads to a maximum prompt length of 2k tokens, with an average of 900 tokens per prompt. Next, we consider the 375M pruned model from Table~\ref{table-prune-summary}, using the full item description with no summarization. Finally, we study the setup where we use the pruned model, as well as the summarized item descriptions for inference. Under this setup, the average prompt length reduces to 220 tokens, a 4x reduction in average prompt length. We report the throughput we can achieve per GPU in Table~\ref{table-offline-bench} for the cases discussed here.

\begin{table}[t]
\caption{Throughput (number of items scored per second per H100 GPU) in an offline setting. The original refers to using the 600M model with full length item description.  Pruned refer to using the 375M pruned model with full length item descriptions. Pruned and Summarized refers to using the 375M model with summarized item descriptions.}
\label{table-offline-bench}
\vskip 0.15in
\begin{center}
\begin{small}
\begin{sc}
\begin{tabular}{lc}
\toprule
Setup & Throughput \\
& (Items/sec/GPU) \\
\midrule
Original & 330 \\
Pruned & 420 \\
Pruned and Summarized &  1530 \\
\bottomrule
\end{tabular}
\end{sc}
\end{small}
\end{center}
\vskip -0.1in
\end{table}

We observe that by using a smaller model (375M parameters instead of 600M), we achieve a 27\% throughput lift (from the engine perspective). Combining pruning and context compression, we see that we achieve a 4.6x throughput lift, while losing less than 2\% of the model's quality (see Table~\ref{table-prune-summary}). 

\subsection{Serving Engine Optimization}
Our semantic search workload differs from standard (e.g., chatbot) workloads in several aspects; First, our workload only includes a prefill phase as we only need the logits corresponding to the last input token to generate our relevance scores based on~\eqref{eq:rel-score}. Second, we work with extremely high number of requests per second, where, for example, even tokenization can become a major bottleneck. Finally, our prompts follow a specific structure, where all prompts corresponding to a query have a shared prefix that include the system instructions and the query itself. This creates opportunities to improve the system throughput through clever Key-Value (KV) cache management in the attention mechanism. Hence, we optimize our online serving pipeline to improve the throughput in a production environment. Below, we discuss some of the techniques we applied.

\paragraph{Batch tokenization} Tokenization---the conversion from raw text to model tokens---is an often-overlooked part of inference latency. For large-scale production systems like LinkedIn’s SLM ranker, tokenization can become a dominant cost at high RPS or during prefill-heavy batched requests.  To address this bottleneck, we have implemented batch tokenization in the SGLang engine. In particular, we tokenize all text prompts inside a single request in one pass, instead of making several calls to the tokenizer. The tokenizer implementation typically performs vectorized work when encoding many strings at once; invoking it once per request amortizes overhead and reduces per-text tokenization time. The relative tokenization speedup (compared to a sequential setup) is reported in Table~\ref{table-batch-tok}.

\begin{table}[t]
\caption{The effect of batch tokenization in SGLang, compared to sequential tokenization. A higher batch size improves tokenization efficiency.}
\label{table-batch-tok}
\vskip 0.15in
\begin{center}
\begin{scriptsize}
\begin{sc}
\begin{tabular}{lc}
\toprule
Batch size & Tokenization speedup\\
\midrule
1 & 1.0x \\
2 & 1.7x \\
4 & 2.8x \\
8 & 5.2× \\
\bottomrule
\end{tabular}
\end{sc}
\end{scriptsize}
\end{center}
\vskip -0.1in
\end{table}

\paragraph{Removing the decode phase}
Scoring/prefill-only workloads do not require per-token sampling or iterative decoding; they only require the final token's distribution (see~\eqref{eq:rel-score}). Running the full decode and sampling loop imposes unnecessary CPU work and memory traffic. To address this issue, we added a scoring-optimized path that skips sampling/decoding phases altogether and computes only the final token probabilities required for scoring. This happens deterministically (no sampling side effects) and avoids the per-token sampling overhead.
In our experiments, we observe that for a single request (10 prompts) with a 300 token prompt length and using the 600M model, the scoring latency reduces from 33 millisecond to 20 milliseconds when bypassing the decode loop.

\paragraph{Reduce memory synchronization}
Even after removing the decode loop, expensive per-item work remains. In particular, per-input-token probability calculations can cause significant latency, while such distributions are not used in our ranking setup (only the last token probabilities are useful). Such calculations with additional sampling and eager memory copies delay subsequent GPU kernel launches (large gaps between kernels), leading to increased latency. To this end:
\begin{enumerate}
    \item We skip internal per-input-token probability extraction when not required.
    \item We introduce functions to compute (last token) probabilities without sampling.
    \item We optimize memory copy operations between CPU and GPU by replacing many small memory copies with a single vectorized gather on the GPU. We also allow for delaying CPU copy operations so CPU postprocessing overlaps with GPU compute for the next batch.
\end{enumerate}
Taken together, for the 600M model with 300 prompt size, the 99-th percentile latency at 100 RPS (1000 prompts per second) plummeted 92.7\% from 6220 milliseconds to 454 milliseconds, allowing for a 25\% increase in throughput.

\paragraph{Fixing garbage collection}
 Long-running SGLang servers occasionally experience periodic Garbage Collection (GC) stalls---100–300 milliseconds pauses occurring roughly every 1.5 seconds---caused by Python’s generational GC scanning over 750K long-lived objects unnecessarily. These stalls caused us to exceed our p99 latency requirements well before the point of GPU saturation. We fixed this by sending warm up requests to ensure all long-lived data structures are initialized and then invoking Python's gc.freeze() API to exclude all the current objects in memory from future scans.

\paragraph{Amortized prefill optimization}  Relevance scoring is essentially a batch prefill where all prompts share the same prefix. Therefore, all prompts in a batch share the same prefix Key-Value (KV) cache. Although off-the-shelf SGLang prefix caching can be used here to reuse this KV cache, it requires two forward passes---once with the first prompt and then again with the remaining prompts to leverage the prefix KV cache. 

Instead, we perform an ``in-batch prefix caching'' where we leverage the KV cache from the first item in the batch to compute the attention scores for the remaining items in the batch, all in one forward pass. To this end, we intercept the forward pass between the computation of the KV cache and the attention scores to store the KV cache corresponding to the prefix. Then, for any subsequent prompt with the same prefix, the attention score is obtained via a merge of two attention operations:
\begin{enumerate}[topsep=0pt,parsep=0pt,partopsep=0pt]
    \item  All suffix tokens attend to all prefix tokens from the first prompt via dense paged attention.
\item All suffix tokens attend to themselves via regular casual attention.
\end{enumerate}
The merge is performed via the Log-Sum-Exp method commonly used to merge two different sets of attention weights. 

For large batch sizes, our benchmarks confirm that throughput gains are roughly proportional to the ratio of reduced to retained token work. Specifically, when the query tokens are skipped, throughput improves according to
\[
T = 1 + \frac{N_q}{N_i},
\]
where \(N_q\) and \(N_i\) denote the number of query and item tokens, respectively. For example, with a batch size of 50, \(N_q = 50\), and \(N_i = 150\), we observe throughput gains of approximately 33\%.

\subsection{Streaming services optimization}
In addition to optimizing the serving inference engine, we perform several optimizations to the streaming and traffic management systems in our semantic search stack, with the goal of increasing the throughput and decreasing GPU under-utilization.

\paragraph{Item score caching}
In LinkedIn’s job search pipeline, a single query can surface hundreds or thousands of potential jobs, yet only 25 results are rendered per page. Each page navigation generates a fresh request to the search service. While the result set changes, the underlying relevance score for a fixed query–item tuple remains deterministic under the same language model. This property makes the scoring step a prime candidate for caching, which not only saves expensive GPU inference cycles but also increases parallel throughput across the ranking service. 

To enhance the efficiency of search services, a distributed cache backed by Couchbase was integrated into the system architecture. Prior to invoking the language model, the system performs a cache lookup: cache hits obviate GPU computation, whereas misses trigger model inference followed by cache updates. This design enables repeated navigations or similar queries to exploit previously computed results, thereby reducing computational overhead. Empirical evaluation indicates that with a configurable 15 minutes TTL, over 50\% of online scoring requests are satisfied directly from the cache. Consequently, latency reductions were observed across both online and nearline traffic. For online traffic, median latency (p50) decreased by 9.9\% and mean latency by 7.3\%, with tail latencies improving by 8.3\% at p90 and 4.4\% at p99. Nearline traffic exhibited similar gains, including an 11.3\% reduction at p50, a 9.2\% decrease in average latency, and tail latency improvements of 7.5\% at p90 and 1.1\% at p99. These results substantiate the effectiveness of the caching strategy in mitigating latency, alleviating GPU load, and reducing overall infrastructure costs.

\paragraph{Dynamic scoring depth}
Search and ranking workloads in large-scale systems rarely follow a flat profile; instead, they oscillate between peak and off-peak periods. To address this variability, we introduced a Dynamic Ranking Depth controller that leverages a PID-based feedback loop to continuously tune how many job candidates flow into the ranking stage. During low traffic, the controller raises ranking depth to exploit unused serving capacity, enabling richer evaluation and higher-quality results for members.

In production, Dynamic Scoring Depth employs a PID controller to adapt ranking depth in response to traffic conditions. Under peak demand, the controller lowers depth just enough to satisfy throughput and latency SLAs while remaining above the neutral-effect threshold to ensure recall is preserved. In controlled product experiments, this adaptive mechanism reduced the average scoring depth from 250 to 131 at peak, yielding a 48\% reduction in per-query GPU compute and allowing the system to accommodate substantially higher traffic volumes with a fixed GPU fleet of practical size. Conversely, during off-peak periods the controller increases depth up to 1000, enhancing result quality without compromising responsiveness. By embedding control-theoretic principles into ranking infrastructure, the system dynamically balances relevance, scalability, and responsiveness across highly variable interactive search and alert workloads.

\paragraph{Traffic shaping} Traffic shaping techniques have also been introduced to handle latency-insensitive workloads and reduce GPU underutilization. Ranking traffic often exhibits bursty arrival patterns: large waves of queries arrive simultaneously, saturating resources, followed by idle gaps where GPUs sit underloaded. By applying controlled deferral and smoothing requests across these gaps, the system pushes a portion of peak RPS into slightly later windows without violating latency requirements. This evens out GPU scheduling, raises kernel efficiency, and drives higher sustained throughput. Production benchmarks indicate a 25\% increase in effective GPU scoring capacity (1600 to 2000 scoring items per second on one Nvidia H100 GPU, measured before and after enabling traffic shaping), demonstrating the impact of shaping burst-heavy workloads through systems-level scheduling.

\section{Online Deployment}
The final SLM combines model compression, context compression, and SFT to recover performance. To validate our offline results and realize the throughput improvements, we deployed this SLM as the online ranking system in LinkedIn's Semantic Job Search stack (see Section~\ref{sec:sjs-intro}), together with the online and streaming optimizations.

\begin{table}[t]
\caption{Online deployment results. The first column shows the performance of the uncompressed SLM against EBR, while the second  shows the performance of the compressed SLM against EBR. We report ``$-$'' when the results are not statistically significant.}
\label{table:ab_tests}
\vskip 0.15in
\begin{center}
\begin{small}
\begin{sc}
\begin{tabular}{lcc}
\toprule
Metric & SLM v1 & SLM v2\\
\midrule
Poor Match Rate @ 10 & $-19.84\%$ & $-26.69\%$\\
NDCG @ 10 & $+5.54\%$ & $+8.45\%$ \\
Job Sessions & $-$ & $+0.59\%$ \\
Jobs Weekly Active Users & $-$ & $+1.49\%$ \\
Job Dismisses & $-$ & $-7.15\%$ \\
\bottomrule
\end{tabular}
\end{sc}
\end{small}
\end{center}
\vskip -0.1in
\end{table}

Due to the unpruned and uncompressed SLM's inefficiency, we were unable to launch it to a significant volume of LinkedIn's users, so a direct online comparison of quality against the unpruned model was infeasible. The majority user experience was therefore just a retrieval-based system, until we could launch the compressed SLM as a second pass reranker. We present two sets of A/B tests to better understand the optimized SLM's online performance:
\begin{enumerate}[noitemsep,topsep=0pt,parsep=0.5pt,partopsep=0pt]
    \item \textbf{Uncompressed SLM (SLM v1) vs EBR:} These results give us a baseline of how a fine-tuned SLM (a 0.5B model) ranking model improves upon a retrieval-only search solution that uses an LLM (7B) for embedding-based retrieval.
    \item \textbf{Compressed SLM (SLM v2) vs EBR:} These results show how the compressed SLM (0.6B model reduced to 375M parameters) using the compressed input context improves upon the same LLM-based EBR.
\end{enumerate}
Note that we updated the SLM's base model from 0.5B to 0.6B in the second test. This resulted in an offline lift of $2\%$ in NDCG@10, which we traded off for improved model compression, giving us similar offline performance as the v1 SLM model. Nevertheless, the SLMs in the two tests have slightly different architectures due to the v2 model having two more query heads, six more KV heads, and four more transformer layers than the v1.

Our results are presented in Table~\ref{table:ab_tests}. Despite similar offline metrics between SLM v1 and v2, we see improvements across the board. The Poor Match Rate metric is a measure of the False Positive Rate of the model. Together with NDCG@10, it shows that the SLM's relevance has improved (the relevance ground truth is given by the SLM's teacher model). The weekly active users and session metrics capture user engagement, while job dismisses indicate their dissatisfaction with results. On both fronts, the SLM v2 shows statistically significant, positive impact. This improved online performance is likely due to each query head being shared among fewer KV heads, and the presence of additional layers in the newer model.

In terms of throughput, we were able to achieve an online throughput of 2000 items scored per second per GPU, using the 375M pruned model and summarized item descriptions. This more than a 10x improvement from v1 model which did not incorporate any of the optimizations discussed in this paper. Thanks to this increase in throughput, we launched LinkedIn's Semantic Job Search to a large segment of users.

\bibliography{example_paper}
\bibliographystyle{mlsys2025}

\end{document}